\documentclass[twocolumn,aps,nofootinbib]{revtex4}
\usepackage{amsfonts, amssymb}
\usepackage{graphicx, epsfig,bm}
\usepackage{color}

\newcommand{\be}{\begin{equation}}
\newcommand{\ee}{\end{equation}}
\newcommand{\bea}{\begin{eqnarray}}
\newcommand{\eea}{\end{eqnarray}}

\def\fun#1#2{\lower3.6pt\vbox{\baselineskip0pt\lineskip.9pt
        \ialign{$\mathsurround=0pt#1\hfill##\hfil$\crcr#2\crcr\sim\crcr}}}

\newcommand\lsim{\mathrel{\rlap{\lower4pt\hbox{\hskip1pt$\sim$}}
    \raise1pt\hbox{$<$}}}
\newcommand\gsim{\mathrel{\rlap{\lower4pt\hbox{\hskip1pt$\sim$}}
    \raise1pt\hbox{$>$}}}

\def\dslash{\not{\hbox{\kern-2pt $\partial$}}}
\def\Dslash{\not{\hbox{\kern-4pt $D$}}}
\def\Oslash{\not{\hbox{\kern-4pt $O$}}}
\def\Qslash{\not{\hbox{\kern-4pt $Q$}}}
\def\pslash{\not{\hbox{\kern-2.3pt $p$}}}
\def\kslash{\not{\hbox{\kern-2.3pt $k$}}}
\def\qslash{\not{\hbox{\kern-2.3pt $q$}}}

 \newtoks\slashfraction
 \slashfraction={.13}
 \def\slash#1{\setbox0\hbox{$ #1 $}
 \setbox0\hbox to \the\slashfraction\wd0{\hss \box0}/\box0 }

\def\ee{\end{equation}}
\def\be{\begin{equation}}

\begin{document}
\setlength{\unitlength}{1mm}
\title{No Evidence for Dark Energy Dynamics\\from a Global Analysis of Cosmological Data}

\author{Paolo Serra\footnote{pserra@uci.edu}$^1$, Asantha Cooray$^1$, Daniel E. Holz$^2$, Alessandro Melchiorri$^{1,3}$, Stefania Pandolfi$^{1,4}$, Devdeep Sarkar$^{1,5}$}
\affiliation{$^1$Center for Cosmology, Department of Physics and Astronomy,
University of California, Irvine, CA 92697}
\affiliation{$^2$Theoretical Division, Los Alamos National Laboratory, Los Alamos, NM 87545}
\affiliation{$^3$Physics Department and Sezione INFN, University of Rome,
``La Sapienza,'' P.le Aldo Moro 2, 00185 Rome, Italy}
\affiliation{$^4$Physics Department and International Centre for Relativistic Astrophysics, University of Rome, ``La Sapienza,'' P.le Aldo Moro 2, 00185 Rome, Italy}

\affiliation{$^5$Physics Department, University of Michigan, Ann Arbor, MI 48109}

\date{\today}%

\begin{abstract}
{We use a variant of principal component analysis to investigate the possible
temporal evolution of the dark energy equation of state, $w(z)$. We constrain
$w(z)$ in multiple redshift bins, utilizing the most recent data from Type Ia
supernovae, the cosmic microwave background, baryon acoustic oscillations, the
integrated Sachs-Wolfe effect, galaxy clustering, and weak lensing data. Unlike
other recent analyses, we find no significant evidence for evolving dark energy;
the data remains completely consistent with a cosmological constant. We also
study the extent to which the time-evolution of the equation of state would be
constrained by a combination of current- and future-generation surveys, such as
Planck and the Joint Dark Energy Mission.}
\end{abstract}

\maketitle

\section{Introduction}
One of the defining challenges for modern cosmology is 
understanding the physical mechanism responsible for the accelerating expansion
of the Universe~\cite{Riess:1998cb,Perlmutter:1998np,DEFT}. The origin of the
cosmic acceleration can be due to a new source of stress-energy, called ``dark energy'', a
modified theory of gravity, or some mixture of both~\cite{Uzan:2006mf,
Copeland2008}. Careful measurement of the expansion history of the Universe as a
function of cosmic epoch is required to elucidate the source of the
acceleration. In particular, existing data already allows direct exploration of
possible time-variation of the dark energy equation of state.

While several recent papers have investigated the possibility of constraining the
temporal evolution of dark energy (see, e.g.,~\cite{latest}), here we present an
analysis improving and/or complementing existing work in two
ways: first, we incorporate important recent data releases, including
Type Ia supernovae samples (``Constitution'' and ``Union'' datasets) and baryon
acoustic oscillation
data (SDSS Data Release 7). This new data provide significant improvements in
the dark energy constraints. Second, we utilize principal component analysis
techniques to constrain the dark energy in a model independent manner, leading
to more robust and unbiased constraints.

In the absence of a well-defined and theoretically motivated model for dark
energy, it is generally assumed that the dark energy equation of state (the ratio of
pressure to energy density) evolves with redshift with an arbitrary functional
form. Common parameterizations include a linear variation,
$w(z)=w_0+w_zz$~\cite{Cooray1999}, or an evolution that asymptotes to a constant
$w$ at high redshift,
$w(z)=w_0+w_az/(1+z)$~\cite{Chevallier2001,Linder2003}. However, given our
complete ignorance of the underlying physical processes, it is advisable to
approach our analysis of dark energy with a minimum of assumptions. Fixing an ad
hoc two parameter form could lead to bias in our inference of the dark energy
properties.

In this paper we measure the evolution history of the dark energy using a
flexible and almost completely model independent 
approach, based on a variant of the principal component analysis (PCA)
introduced in~\cite{Huterer2003}.
We determine the equation of state parameter, $w(z)$, in five
uncorrelated redshift bins,
following the analysis presented in~\cite{HutererCooray2005,Sarkar2007,Zhao2007,Sullivan07}. To be 
conservative, we begin by using data only from geometric probes of dark energy,
namely the cosmic microwave background radiation (CMB), Type Ia supernovae (SNe)
and baryon acoustic oscillation data (BAO). We perform a full likelihood analysis
using the Markov Chain Monte Carlo approach~\cite{Lewis2002}.  We then consider
constraints on $w(z)$ from a larger combination of datasets, including probes of
the growth of cosmological perturbations, such as large scale structure
(LSS) data.  An important consideration for such an analysis is to properly take
into account dark energy perturbations, and we make use of the prescription
introduced in~\cite{Feng2008}.  We also generate mock datasets for
future experiments, such as the Joint Dark Energy Mission (JDEM) and Planck, to
see how much they improve the constraints.

The paper is organized as follows: in the next section we describe our methods
and the data used in our analysis; in Sec.~III we present
our results, and in Sec.~IV we summarize and conclude.

\begin{figure}[!t]
\begin{center}
\begin{tabular}{cc}
 \resizebox{80mm}{!}{\includegraphics{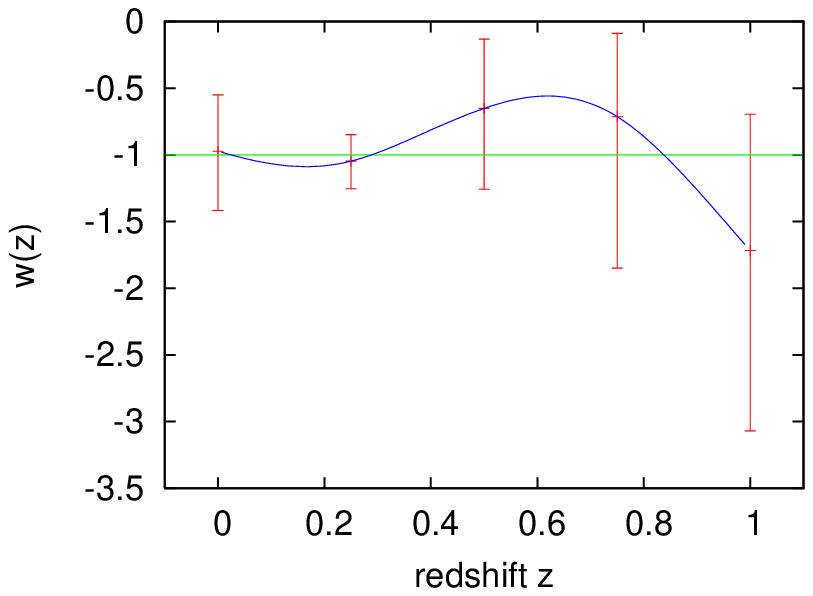}}\\
\resizebox{80mm}{!}{\includegraphics{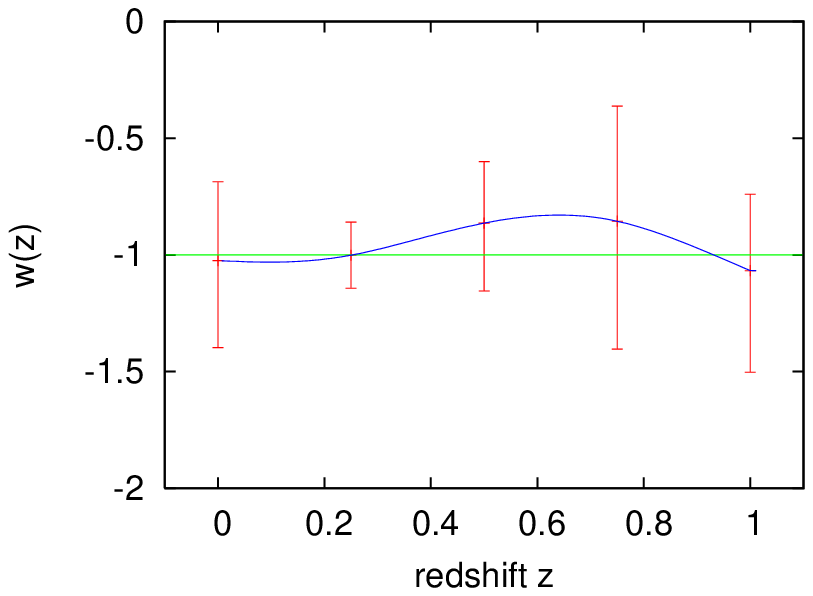}}\\
\resizebox{80mm}{!}{\includegraphics{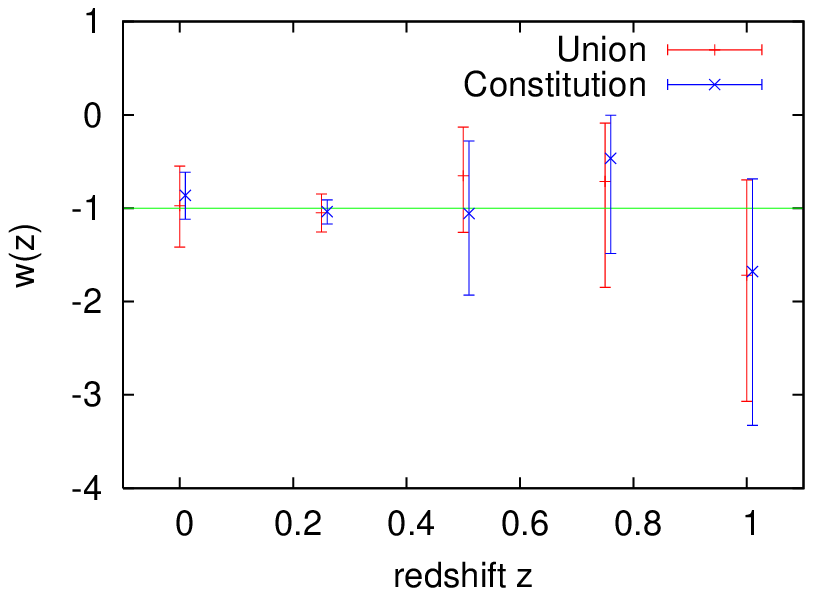}}\\
 \end{tabular}
\caption{Uncorrelated constraints on the dark equation of state parameters using a ``geometric'' dataset given by
WMAP+UNION+BAO (upper panel), and a ``combined'' dataset given by
WMAP+UNION+BAO+WL+ISW+LSS (middle panel); error bars are at
$2\sigma$. The blue line is the reconstructed $w(z)$ using a cubic spline
interpolation between the nodes. Also shown is a comparison between
WMAP+UNION+BAO and WMAP+Constitution+BAO (lower panel); the points for the
Constitution dataset have been slightly shifted to facilitate comparison
between the two cases: we find no significant difference between UNION and
Constitution.} 
\label{fig.1}
\end{center}
\end{figure}

\section{Analysis}
The method we use to constrain the dark energy evolution is based on a modified
version of the publicly available Markov Chain Monte Carlo package CosmoMC
\cite{Lewis2002}, with a convergence diagnostics based on the Gelman-Rubin
criterion~\cite{gelman}. We consider a flat cosmological model described by the
following set of parameters:
\begin{equation}
 \label{parameter}
      \{w_i,\omega_{b}, \omega_{c},
      \Theta_{s}, \tau,  n_{s}, \log[10^{10}A_{s}] \}~,
\end{equation}
where $\omega_{b}$ ($\equiv\Omega_{b}h^{2}$) and $\omega_{c}$ ($\equiv\Omega_{c}h^{2}$) are the physical baryon and cold dark matter densities relative to the critical density,
$\Theta_{s}$ is the ratio of the sound horizon to the angular
diameter distance at decoupling, $\tau$ is the optical depth to re-ionization,
and $A_{s}$ and $n_s$ are the amplitude of the primordial spectrum and the
spectral index, respectively. 

As discussed above, we bin the dark energy equation of state in five redshift
bins, $w_i(z)\,(i=1,2,..5)$, representing the value at five redshifts, 
$z_i\in{[0.0,0.25,0.50,0.75,1.0]}$. We have explicitly verified that the use of
more than five bins does not significantly improve the dark energy constraints.
We need $w(z)$ to be a smooth, continuous function, since we evaluate $w'(z)$
in calculating the DE perturbations (and their evolution with
redshift). We thus utilize a cubic spline interpolation to
determine values of $w(z)$ at redshifts in between the values $z_i$.

For $z>1$ we fix the equation of state parameter at its $z=1$ value, since we
find that current data place only weak
constraints on $w(z)$ for $z>1$. To summarize, our parameterization is given by:
\begin{equation} \label{paraw} w(z)= \left\{
    \begin{array}{ll}
          w(z=1),&  \hbox{$z> 1$;} \\
      \hbox{  $w_i$}, & \hbox{$z\leq z_{max}, z\in \{z_i\}$;} \\
      \hbox{ spline}, & \hbox{$z\leq z_{max}, z\notin \{z_i\}$.}
    \end{array}
\right. \end{equation}

\begin{table*}
\caption{Mean values and marginalized $68 \%$ confidence levels for the
cosmological parameters. The set of ${w(z)}_i$ represent the measured values of
the dark energy equation of state in uncorrelated redshift bins.}
\begin{center}
\begin{tabular}{|c|c|c|c|c|}
\hline
Parameter &WMAP+UNION+BAO & WMAP+Constitution+BAO &all dataset & future datasets \\
\hline
$\Omega_bh^2$&$0.02281\pm0.00057$&$0.02278\pm0.00058$&$0.02304\pm0.00056$ &$0.02270\pm0.00015$\\
$\Omega_{\rm c}h^2$&$0.1128\pm0.0059$&$0.1144\pm0.0060$&$0.1127\pm0.0018$&$0.1100\pm0.0012$ \\
$\Omega_{\Lambda}$&$0.728\pm 0.018$&$0.715\pm 0.017$&$0.728\pm0.016$&$0.751\pm0.008$\\
$n_s$&$0.964\pm0.014$&$0.963\pm0.014$&$0.971\pm0.014$ &$0.962\pm0.004$\\
$\tau$&$0.085\pm0.017$&$0.084\pm0.016$&$0.088\pm0.017$ &$0.084\pm0.05$\\
$\Delta^2_R$&$(2.40\pm0.10)\cdot10^{-9}$&$(2.40\pm0.10)\cdot10^{-9}$&$(2.40\pm0.10)\cdot10^{-9}$&$(2.40\pm0.10)\cdot10^{-9}$\\
$w(z=1.7)$& $--$&$--$&$--$&$-1.55_{-0.44}^{+0.46}$\\
$w(z=1)$& $-1.72^{+0.73}_{-0.81}$&$-1.68^{+0.73}_{-0.85}$&$-1.07_{-0.20}^{+0.21}$&$-1.03\pm0.10$\\
$w(z=0.75)$& $-0.71^{+0.44}_{-0.47}$&$-0.47^{+0.34}_{-0.33}$&$-0.86^{+0.025}_{-0.26}$ &$-0.98\pm0.08$\\
$w(z=0.5)$& $-0.65^{+0.29}_{-0.30}$&$-1.06^{+0.41}_{-0.40}$&$-0.86\pm0.14$&$-1.00\pm0.05$ \\
$w(z=0.25)$& $-1.05\pm0.10$&$-1.04\pm0.07$&$-1.00\pm0.07$&$-1.00\pm0.02$\\
$w(z=0)$& $-0.97\pm0.22$&$-0.86\pm0.13$&$-1.02_{-0.18}^{+0.17}$&$-0.99\pm0.05$ \\
\hline
$\sigma_8$&$0.814\pm0.055$&$0.815\pm0.057$& $0.810\pm 0.024$ &$0.811\pm0.012$\\
$\Omega_m$&$0.272\pm0.018$&$0.285\pm0.017$&$0.272\pm0.016$ &$0.249\pm0.008$\\
$H_0$&$70.7\pm2.0$&$69.4\pm1.7$&$70.8\pm2.0$ &$73.1\pm1.0$\\
$z_{reion}$&$10.8\pm1.4$&$10.8\pm1.4$& $11.0\pm1.5$ &$10.7\pm0.4$\\
$t_0$&$13.65\pm0.14$&$13.67\pm0.15$&$13.67\pm0.13$&$13.60\pm0.06$\\
\hline

 \end{tabular}
 \label{table:1}
 \end{center}
 \end{table*}

When fitting to the temporal evolution of the dark energy equation of state
using cosmological measurements that are sensitive to density perturbations,
such as LSS or weak lensing, one must take into account the presence of dark
energy perturbations. To this end, we make use of a modified version of the
publicly available code CAMB~\cite{Lewis:1999bs}, with perturbations calculated
following the prescription introduced by~\cite{Feng2008}. This method
implements a Parameterized Post-Friedmann (PPF) prescription for the dark energy
perturbations following~\cite{Hu2007,Hu2008}.

Moreover, the dark energy equation of state parameters ${\bf w}={w_i}$ are
correlated; we follow~\cite{HutererCooray2005,Sarkar2007} to determine
uncorrelated estimates of the dark energy parameters. We calculate the
covariance matrix $\textbf{C}=(w_i-\langle w_i\rangle)(w_j-\langle
w_j\rangle)^T\equiv \langle {\bf w} {\bf w}^T\rangle-\langle {\bf
w}\rangle\langle {\bf w}^T\rangle$, using CosmoMC; we then diagonalize
the resulting Fisher matrix ${\bf F}\equiv {\bf C}^{-1}$, which can be written
as ${\bf F}={\bf O^T\,\Lambda\,O}$, where ${\bf \Lambda}$ is the diagonalized
inverse covariance of the transformed bins. The vector of uncorrelated dark
energy parameters, ${\bf q}$, is then obtained from ${\bf q}={\bf
Ow}$. If we now define $\tilde{{\bf W}}$ so that $\tilde{{\bf W}^T}\tilde{{\bf
W}}={\bf F}$ then, as emphasized by~\cite{Hamilton}, there are infinitely
many choices for the matrix ${\tilde{\bf W}}$;
following~\cite{HutererCooray2005}, we write the weight transformation matrix as
$\tilde{{\bf W}}={\bf O^T\Lambda^{\frac{1}{2}}O}$ where the rows are summed such
that the weights from each band add up to unity, and we apply this
transformation matrix to obtain our uncorrelated estimates of dark energy
parameters. Our first analysis considers constraints from ``geometric'' data:
CMB, Type Ia SN luminosity distances, and BAO data. We subsequently include
datasets that probe the growth of cosmic structures, incorporating weak
lensing, as well as integrated Sachs-Wolfe measurements through cross-correlations between CMB and galaxy survey data.
We include the latter datasets
separately, since our understanding of the cosmic clustering in dark energy
models still suffers from several limitations. These LSS uncertainties are mainly related to our poor
understanding of both the bias between galaxies and matter fluctuations (with a
possible scale dependence of the bias itself,
see~\cite{McDonald2006,Smith2006,Joudaki2009}) and non-linearities at small
redshifts (see~\cite{Dunkley:2008ie,Hamann2008}).
For the CMB, we use data and likelihood
code from the WMAP team's 5-year release~\cite{Dunkley:2008ie} (both temperature
TT and polarization TE; we will refer to this analysis as WMAP5). In this
respect, our approach is more extensive than that in~\cite{latest} and other
recent studies, since we fully consider the CMB dataset
instead of simply using the constraint on the $\theta$ parameter from the
analysis of~\cite{Dunkley:2008ie}.  This constraint is model dependent
(see, e.g.,~\cite{corasaniti}), and changes with dark energy parameterizations.

Supernova data come from the Union data set (UNION) produced by the Supernova
Cosmology Project~\cite{Kowalski:2008ez}; however, to check the consistency of
our results, we also used the recently released Constitution dataset
(Constitution)~\cite{Hicken2009} which, with 397 Type Ia supernovae, is the
largest sample to date. We also used the latest SDSS release (DR7) BAO distance
scale~\cite{Reid2009,Percival2009}: at $z=0.275$ we have
$r_s(z_d)/D_V(0.275)=0.1390\pm0.0037$ (where $r_s(z_d)$ is the comoving sound
horizon at the baryon drag epoch,
$D_V\equiv[(1+z)^2D_A^2cz/H(z)]^{\frac{1}{3}}$, $D_A(z)$ is the angular diameter
distance and $H(z)$ is the Hubble parameter) and the ratio of distances
$D_V(0.35)/D_V(0.20)=1.736\pm0.065$. Weak lensing (WL) data are taken from
CFHTLS~\cite{Fu2007} and we use the weak lensing module provided
in~\cite{Massey,Lesgo}, with some modifications to assess the likelihood in
terms of the variance of the aperture mass (Eq.~5 of~\cite{Fu2007}) with
the full covariance matrix~\cite{Kilbinger}.  The cross-correlation between CMB
and galaxy survey data is employed using the public code at~\cite{Howebsite}. We
modify it to take into account the temporal evolution
of the dark energy equation of state, since the code only considers $w$CDM
cosmologies. We refer to~\cite{Ho1,Ho2} for a description of both the
methodology and the datasets used. Finally, we use the recent value of the Hubble constant from the SHOES
(Supernovae and $H_0$ for the Equation of State)
program,
$H_{0}=74.2\pm3.6$~km~s$^{-1}$~Mpc$^{-1}(1\sigma)$~\cite{Reiss2009}, which
updates the value obtained
from the Hubble Key Project~\cite{Hubble}. We also incorporate baryon density
information from Big Bang Nucleosynthesis $\Omega_{b}h^{2}=0.022\pm0.002$
($1\sigma$)~\cite{Burles}, as well as a top-hat prior on the age of the
Universe, $10\mbox{ Gyr}<t_0< 20\mbox{ Gyr}$. 

\begin{figure}[!t]
\begin{center}
\includegraphics[scale=0.9]{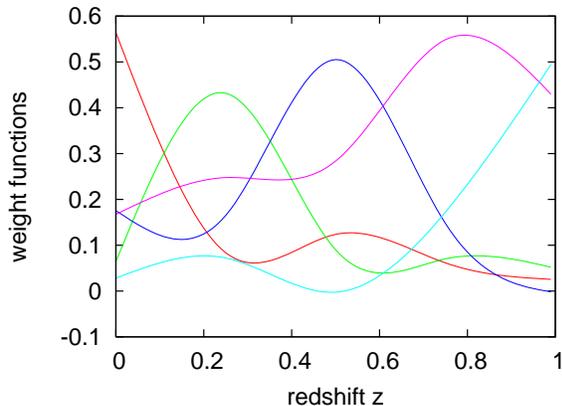}
 \caption{Weight functions for each of the uncorrelated bins, for the case
 WMAP+UNION+BAO+WL+ISW+LSS.}
\label{fig.2}
\end{center}
\end{figure}

\section{Results}
In Table~1 we show the mean values and marginalized $68\%$ confidence level limits for the
cosmological parameters considered in this analysis for the WMAP+SNe(UNION)+BAO
and WMAP+SNe(Constitution)+BAO datasets. We also consider a ``global'' dataset:
WMAP+SNe+BAO+CFHTLS+CMB+WL+ISW+ LSS. The
$w_i(z)\,(i=1,2,..5)$ entries refer to the uncorrelated
values of the dark energy equation of state parameters. All values are
compatible with a cosmological constant ($w=-1$) at the $2\sigma$ level. As we can
see from Table~1 and from Figure~1, there is no discrepancy between the Union and
Constitution datasets; moreover, the addition of cosmological probes of cosmic
clustering noticeably reduces the uncertainty in the determination of the dark
energy parameters, especially at high redshifts.

\begin{figure}[!b]
\begin{center}
\includegraphics[scale=0.9]{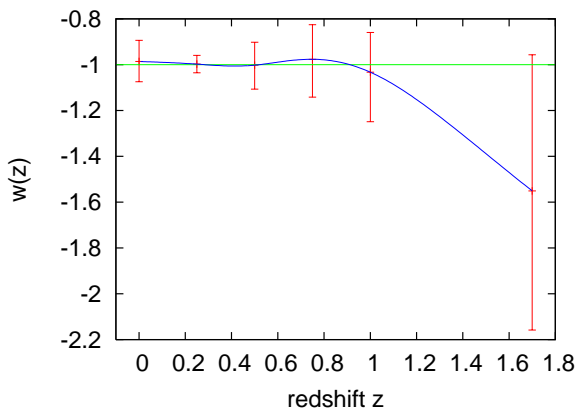}
 \caption{Uncorrelated constraints on the dark energy equation of state parameters, for mock
datasets from Planck and JDEM; error bars are at $2\sigma$.}
\label{fig.3}
\end{center}
\end{figure}


To reinforce our conclusions, we also created several mock datasets for upcoming
and future SN, BAO, and CMB experiments. The quality of future datasets allows
us to constrain the dark energy evolution beyond redshift $z=1$. We thus
consider an additional bin at $z=1.7$, with a similar constraint:
$w(z>1.7)=w(z=1.7)$. We consider a mock catalog of 2,298 SNe, with 300 SNe
uniformly distributed out to z = 0.1, as expected from ground-based low redshift
samples, and an additional 1998 SNe binned in 32 redshift bins in the range $0.1
< z < 1.7$, as expected from JDEM or similar future surveys~\citep{kim:04}. The
error in the distance modulus for each SN is given by the intrinsic
error, $\sigma_{\mbox{int}}=0.1\,\mbox{mag}$. In generating the SN catalog, we do
not include the effect of gravitational lensing, as these are expected to be
small~\citep{sarkar:08}. In addition, we use
a mock catalog of 13 BAO estimates, including 2 BAO estimates at z = 0.2 and
z = 0.35, with 6\% and 4.7\% uncertainties (in $D_V$), respectively, 4 BAO
constraints at $z=[0.6,0.8,1.0,1.2]$ with corresponding fiducial survey
precisions (in $D_V$) of $[1.9,1.5,1.0,0.9]\%$ (V5N5 from~\cite{Seo03}), and 7
BAO estimates with precision $[0.36,0.33,0.34,0.33,0.31,0.33, 0.32]\%$
from $z=1.05$ to $z=1.65$ in steps of 0.1~\cite{private}.

We simulate Planck data using a fiducial $\Lambda$CDM model, with the
best fit parameters from WMAP5, and noise properties consistent with a combination
of the Planck $100$--$143$--$217$ GHz channels of the HFI~\cite{Planck2009}, and
fitting for temperature and polarization using the full-sky likelihood function given
in~\cite{LewisPlanck}.  In addition, we use the same priors on the Hubble parameter
and on the baryon density as considered above.  As can be seen
from Table~1 and Figure~3, future data will reduce the uncertainties in ${w_i}$ by a
factor of at least $2$, with the relative uncertainty below $10\%$ in all but
the last bin (at $z=1.7$).

\section{Conclusions}

One of the main tasks for present and future dark energy surveys is to determine whether
or not the dark energy density is evolving with time.  We have performed a
global analysis of the latest cosmological datasets, and have constrained the
dark energy equation of state using a very flexible and almost model independent
parameterization. We determine the equation of state $w(z)$ in five independent
redshift bins, incorporating the effects of dark energy perturbations. We find
no evidence for a temporal evolution of dark energy---{\em the data is completely consistent
with a cosmological constant}. This agrees with most previous
results, but significantly improves the overall
constraints~\cite{HutererCooray2005,Sarkar2007,Zhao2007,Sullivan07}.

Bayesian evidence models strongly suggests that the dark
energy is a cosmological constant, given that the cosmological constant remains
a very good fit to the data as the number of dark energy parameters increases
(see e.g.~\cite{serra07} and references therein). We show that future 
experiments, such as Planck or JDEM, will be able to reduce the
uncertainty on $w(z)$ to less than $10\%$ in multiple redshift bins,
thereby mapping any temporal evolution of dark energy with
high precision. With this data it will be possible to measure
the temporal derivative of the equation of state parameters, ${dw}/{dz}$,
useful in discriminating between two broad classes of ``thawing'' and
``freezing'' models~\cite{Caldwell2009}.

\smallskip
\noindent
{\it Note\/}: As we were completing this paper we became aware of the work
reported in~\cite{Zhao2009}, which considers a similar analysis of cosmological
data to constrain $w(z)$. While those authors find weak evidence for evolution
of the EOS, we find no such evidence. The two analyses differ in the way $w(z)$
is interpolated (we use a spline, while they employ a tanh function), as well as
different calculations of the effects of DE perturbations. Furthermore, we
analyze different datasets; in this paper we have utilized both the
latest BAO measurements~\cite{Reid2009,Percival2009}, and the latest value of
the Hubble constant from the SHOES program~\cite{Reiss2009}.

 \smallskip
PS acknowledges Alexandre Amblard for useful discussions and Shirley Ho for help
with the ISW likelihood code. This research was funded by NSF CAREER AST-0605427
and by LANL IGPP-08-505.


\end{document}